\documentclass[aps,prl,preprint,groupedaddress,amsmath,amssymb]{revtex4}

\usepackage{graphicx}
\usepackage{dcolumn}
\usepackage{bm}
\usepackage{lineno}
\usepackage[a4paper]{geometry}
\usepackage{setspace}
\usepackage{color, soul}

\begin{document}
\title{Effect of carbon doping on the structure and superconductivity in AlB$_{2}$-type (Mo$_{0.96}$Ti$_{0.04}$)$_{0.8}$B$_{2}$}

\date{\today}
\author{Wuzhang Yang$^{1,2,3}$}
\author{Guorui Xiao$^{2,3,4}$}
\author{Qinqing Zhu$^{2,3,1}$}
\author{Shijie Song$^{4}$}
\author{Guang-Han Cao$^{4}$}
\author{Zhi Ren$^{2,3}$\footnote{Email: renzhi@westlake.edu.cn}}

\affiliation{$^{1}$Department of Physics, Fudan University, Shanghai 200433, P. R. China}
\affiliation{$^{2}$Department of Physics, School of Science, Westlake University, 18 Shilongshan Road, Hangzhou 310024, P. R. China}
\affiliation{$^{3}$Institute of Natural Sciences, Westlake Institute for Advanced Study, 18 Shilongshan Road, Hangzhou 310024, P. R. China}
\affiliation{$^{4}$Department of Physics, Zhejiang University, Hangzhou 310027, P. R. China}

\begin{abstract}
We report the effect of carbon doping in Ti-stabilized nonstoichiometric molybdenum diboride (Mo$_{0.96}$Ti$_{0.04}$)$_{0.8}$B$_{2}$, which exhibits bulk superconductivity below $T_{\rm c}$ = 7.0 K.
It is found that (Mo$_{0.96}$Ti$_{0.04}$)$_{0.8}$(B$_{1-x}$C$_{x}$)$_{2}$ maintains the AlB$_{2}$-type phase with a uniform elemental distribution for $x$ = 0.12 and 0.16.
The substitution of carbon for boron leads to a slight increase in $a$-axis, a remarkable reduction in $c$-axis, the formation of planar defects along the (100) crystallographic planes, and a shift of the B 1$s$ peaks towards higher binding energies.
Contrary to (Mo$_{0.96}$Ti$_{0.04}$)$_{0.8}$B$_{2}$, however, no superconductivity is observed down to 1.8 K for the C-doped samples, which is ascribed to the electron filling of boron $\pi$ bands resulting from the carbon doping.\\
\end{abstract}

\maketitle
\maketitle
\section{1. Introduction}
Transition metal diborides (TMDBs) have attracted a lot of attention over the past few decades due to their diverse properties \cite{TMD1,TMD2,TMD3}, including high hardness and shear strength \cite{hardness1,hardness2,hardness3,hardness4}, excellent corrosion resistance \cite{corrosion1,corrosion2}, superior catalytic performance \cite{catalytic1,catalytic2}, notrivial band topology \cite{topology1,topology2}, high electrical conductivity and even superconductivity \cite{SC1,SC2,SC3,SC4,SC5}.
Among existing TMDBs, molybdenum diboride is unique in that it exists in both hexagonal AlB$_{2}$-type and rhombohedral structures \cite{MoB2}.
The former is built up by planar transition metal and boron layers stacked alternatively along the $c$-axis, and the latter can be derived from the former by puckering half of the boron layers.
Although both phases are not superconducting, superconductivity can be induced by either doping at the transition metal site or applying high pressure. In particular, the $T_{\rm c}$ of AlB$_{2}$-type MoB$_{2}$ reaches 32 K at $\sim$110 GPa \cite{MoB2pressureSC}, comparable to that of isostructural MgB$_{2}$ \cite{MgB2}. At ambient pressure, however, MoB$_{2}$ crystallizes in the rhombohedral structure.
Instead, partial substitution of Mo by Zr or Sc is needed to stabilize the AlB$_{2}$-type phase in metal-deficient (Mo$_{x}$$X$$_{1-x}$)$_{1-\delta}$B$_{2}$ ($X$ = Zr and Sc), which exhibits bulk superconductivity up to 8.3 K \cite{SC3,SC5}.
In addition to Zr, preliminary results are also reported for the group IVB dopants Ti and Hf \cite{SC3}.

Carbon is known to be the most effective dopant for enhancing the upper critical field of MgB$_{2}$ \cite{MgB2C1,MgB2C2}.
Single crystal results indicate that carbon can substitute up to $\sim$15\% of boron in MgB$_{2}$, which leads to a monotonic decreases in the $a$-axis while has little effect on the $c$-axis \cite{MgB2C3,MgB2C4}.
With increasing carbon content, the $T_{\rm c}$ of Mg(B$_{1-x}$C$_{x}$)$_{2}$ decreases dramatically and is no longer detectable for $x$ $>$ 0.125.
Despite the suppression of $T_{\rm c}$, the zero-temperature upper critical field $B_{\rm c2}$(0) is almost doubled with increasing $x$ from 0 to $\sim$0.04 \cite{MgB2C1,MgB2C2}.
To further optimize the high field performance, codoping of Mg(B$_{1-x}$C$_{x}$)$_{2}$ with Ti was also attempted \cite{MgB2TiC}.
It turns out that the carbon still replaces boron in the MgB$_{2}$ phase while Ti precipitates out as either TiB or TiB$_{2}$ in the intra-granular region.
However, neither carbon doping nor its effect on superconductivity in TMDBs has been investigated to date.

In this paper, we study the structure and physical properties of (Mo$_{0.96}$Ti$_{0.04}$)$_{0.8}$(B$_{1-x}$C$_{x}$)$_{2}$.
Structural analysis indicates the formation of an AlB$_{2}$-type phase for $x$ = 0, 0.12 and 0.16.
The carbon dopant is distributed uniformly in the lattice and induces changes in the lattice parameters, microstructure, and boron binding energies.
Physical property measurements show that (Mo$_{0.96}$Ti$_{0.04}$)$_{0.8}$B$_{2}$ exhibits bulk superconductivity below 7.0 K while the C-doped samples remain normal down to 1.8 K, the reason for which is discussed.

\section{2. Experimental section}
Polycrystalline (Mo$_{0.96}$Ti$_{0.04}$)$_{0.8}$(B$_{1-x}$C$_{x}$)$_{2}$ samples with $x$ = 0, 0.04, 0.08, 0.12 and 0.16 were synthesized by the arc melting method as described previously \cite{SC3,SC5}.
Stoichiometric amounts of high purity Mo (99.99\%), Ti (99.99\%), B (99.99\%) and C (99.99\%) powders were weighed, mixed thoroughly and pressed into pellets in an argon-filled glovebox.
The pellets were then melted in an arc furnace under high-purity argon atmosphere (99.999\%) with a current of 80 A, which roughly corresponds to a temperature of 2400 $^{\circ}$C. The melts were flipped and remelted at least four times to ensure homogeneity, followed by rapid cooling on a water-chilled copper plate.
The phase purity of resulting samples was checked by powder x-ray diffraction (XRD) measurements in the 2$\theta$ range of 20-80$^{\circ}$ using a Bruker D8 Advance x-ray diffractometer with Cu-K$\alpha$ radiation.
The step size is 0.025$^{\circ}$ and the dwelling time at each step is 0.1 s.
The lattice parameters were determined by a least-squares method.
The morphology and chemical composition were investigated in a Zeiss Supratm 55 Schottky field emission scanning electron microscope (SEM) equipped with an energy dispersive x-ray (EDX) spectrometer.
The microstructure was examined in an FEI Tecnai G2 F20 S-TWIN transmission electron microscope (TEM) operated under an accelerating voltage of 200 kV.
The x-ray photoelectron spectroscopy (XPS) measurements were performed in a ESCALAB Xi+ spectrometer with Al K$\alpha$ x-rays as the excitation source.
The resistivity and specific heat measurements down to 1.8 K were done in a Quantum Design Physical Property Measurement System (PPMS-9 Dynacool),
The resistivity was measured on bar-shaped samples using the standard four-probe method, and the applied current is 1 mA.
The dc magnetization was measured down to 1.8 K in a commercial SQUID magnetometer (MPMS3) with an applied field of 1 mT.

\section{3. Results and discussion}
\noindent\textbf{3.1 X-ray structural analysis}\\
Figure 1(a) shows the XRD patterns for the (Mo$_{0.96}$Ti$_{0.04}$)$_{0.8}$(B$_{1-x}$C$_{x}$)$_{2}$ samples with $x$ up to 0.16.
In line with the previous report \cite{SC3}, the C-free (Mo$_{0.96}$Ti$_{0.04}$)$_{0.8}$B$_{2}$ ($x$ = 0) sample has a dominant hexagonal AlB$_{2}$-type phase ($P$6/$mmm$ space group).
The refined lattice parameters are $a$ = 3.042 {\AA} and $c$ = 3.150 {\AA}, which are smaller than those of (Mo$_{0.96}$Zr$_{0.04}$)$_{0.8}$B$_{2}$ \cite{SC3}.
However, substituting as low as 4\% of B by C destabilizes the AlB$_{2}$-type structure. Indeed, the sample with $x$ = 0.04 contains a significant amount of rhombohedral Mo$_{2}$B$_{5}$-type impurity phase. Note that this phase is rich in boron, which may explain the absence of boron peak in the pattern.
As the carbon content $x$ increases further, the impurity phase is suppressed rapidly and nearly single AlB$_{2}$-type phase is recovered for $x$ = 0.12 and 0.16.
Especially, the absence of diffraction peaks from carbide phases provides compelling evidence that carbon is incorporated into the lattice, which, to our knowledge, is the first observation for AlB$_{2}$-type TMDs.
The $a$- and $c$-axis lengths are found to be 3.048 {\AA} and 3.094 {\AA} for $x$ = 0.12, and 3.050 {\AA} and 3.091 {\AA} for $x$ = 0.16.
Compared with (Mo$_{0.96}$Ti$_{0.04}$)$_{0.8}$B$_{2}$, the C-doped samples have slightly longer $a$-axes and considerably shorter $c$-axes.
Note that carbon has a smaller atomic radius and a larger electronegativity than boron \cite{radius}.
Hence the substitution of carbon for boron is expected to reduce the thickness of boron layers and enhance their attraction with the transition metal layers,
both of which tend to shorten the $c$-axis.
Meanwhile, the transition metal atoms could move apart so that the Coulomb repulsion between them is reduced and consequently the $a$-axis increases.
\begin{figure}
\includegraphics[width=14.6cm]{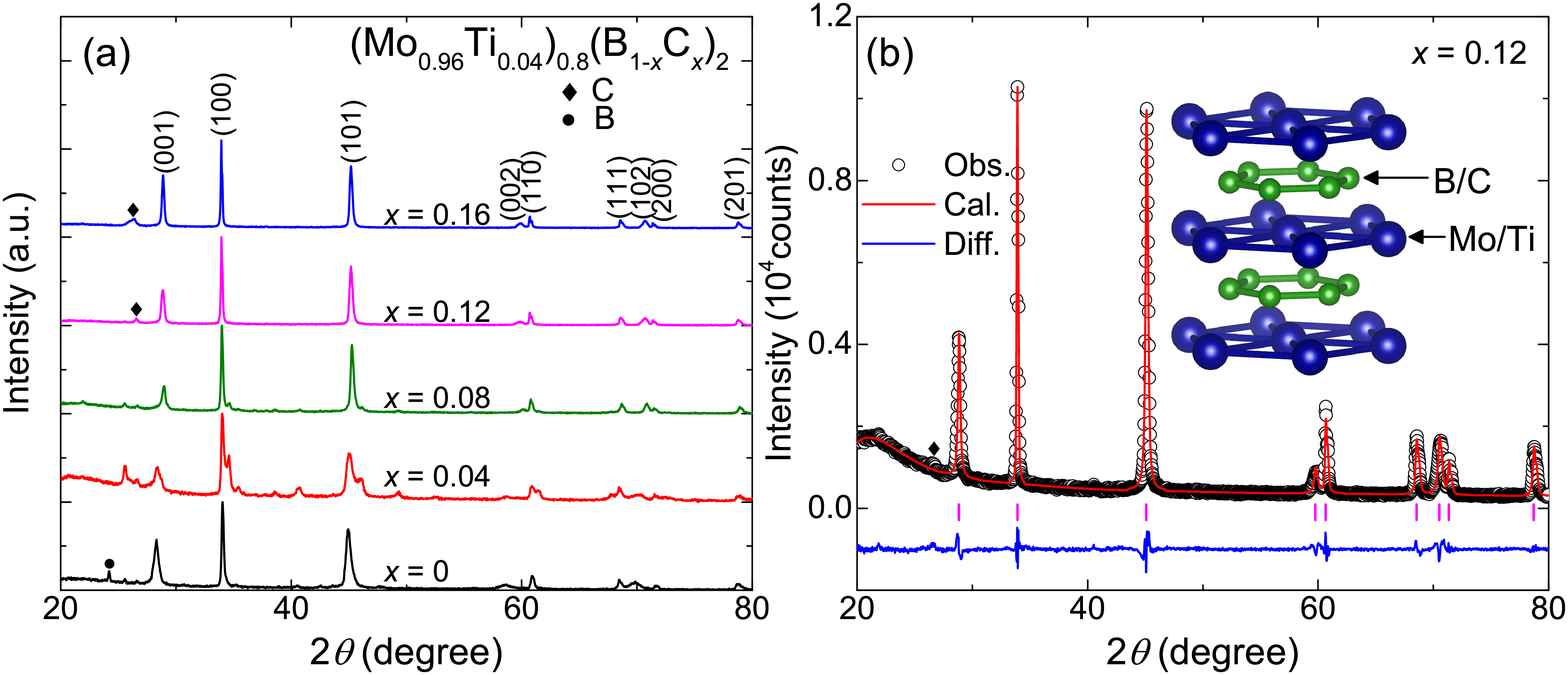}
\caption{
(a) XRD patterns of the (Mo$_{0.96}$Ti$_{0.04}$)$_{0.8}$(B$_{1-x}$C$_{x}$)$_{2}$ samples with 0 $\leq$ $x$ $\leq$ 0.16. The peaks related to different impurities are marked by different symbols. (b) Structural refinement profile for the sample with $x$ = 0.12. Here black circles and red line are the observed (Obs.) and calculated (Cal.) patterns, respectively; the blue solid line represents the difference (Diff.) between them; the magenta ticks represent the expected positions of Bragg reflections.The inset shows a schematic structure of (Mo$_{0.96}$Ti$_{0.04}$)$_{0.8}$(B$_{1-x}$C$_{x}$)$_{2}$.
}
\label{fig1}
\end{figure}

\begin{figure*}
\includegraphics*[width=13.5cm]{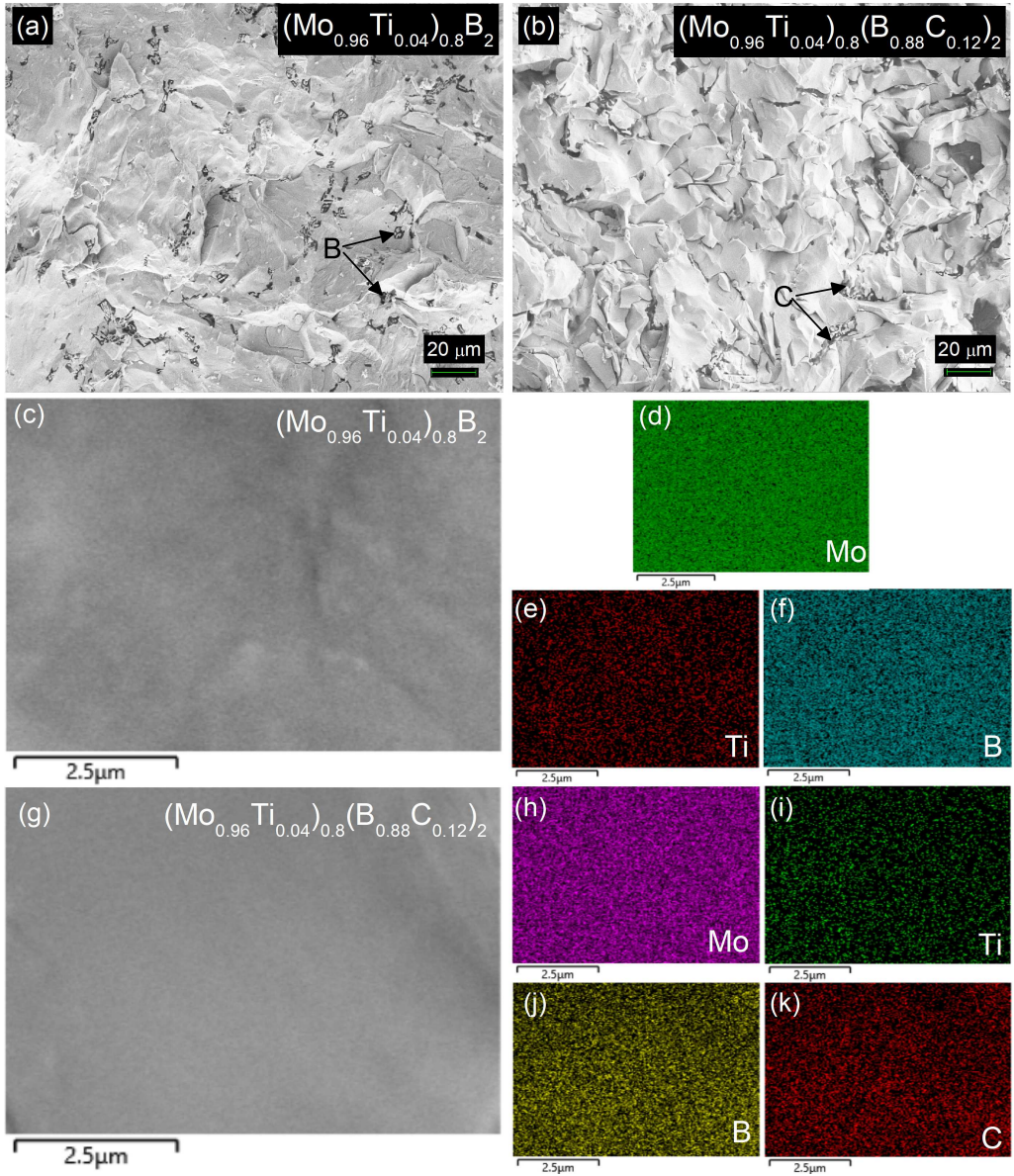}
\caption{
(a, b) SEM images for the (Mo$_{0.96}$Ti$_{0.04}$)$_{0.8}$B$_{2}$ and (Mo$_{0.96}$Ti$_{0.04}$)$_{0.8}$(B$_{0.88}$C$_{0.12}$)$_{2}$ samples, respectively, on a scale bar of 20 $\mu$m.
(c) SEM image for the (Mo$_{0.96}$Ti$_{0.04}$)$_{0.8}$B$_{2}$ sample on a scale bar of 2.5 $\mu$m.
(d-f) EDX elemental mapping results for Mo, Ti, and B, respectively.
(g) SEM iamge for the (Mo$_{0.96}$Ti$_{0.04}$)$_{0.8}$(B$_{0.88}$C$_{0.12}$)$_{2}$ sample on a scale bar of 2.5 $\mu$m. (h-k) EDX elemental mapping results for Mo, Ti, B, and C, respectively.
}
\label{fig2}
\end{figure*}
The structural refinement profile for the sample with $x$ = 0.12 is displayed in Fig. 1(b).
In the unit cell of AlB$_{2}$, the Al and B atoms occupy the (0, 0, 0) and (0,3333, 0.6667, 0.5) sites, respectively.
For (Mo$_{0.96}$Ti$_{0.04}$)$_{0.8}$(B$_{1-x}$C$_{x}$)$_{2}$, the Mo and Ti atoms are set to share the former site while the B and C atoms are set to share the latter one.
As can be seen, all the diffraction peaks can be well fitted based on this structural model, which is corroborated by the small reliability factors ($R_{\rm wp}$ = 6.0\% $R_{\rm p}$ = 4.3\%).
These results confirm the AlB$_{2}$-type structure of (Mo$_{0.96}$Ti$_{0.04}$)$_{0.8}$(B$_{1-x}$C$_{x}$)$_{2}$, which is sketched in the inset of Fig. 1(b).\\

\noindent\textbf{3.2 Morphology, chemical composition and microstructure}\\
Typical SEM images on a scale bar of 20 $\mu$m for the (Mo$_{0.96}$Ti$_{0.04}$)$_{0.8}$B$_{2}$ and (Mo$_{0.96}$Ti$_{0.04}$)$_{0.8}$(B$_{0.88}$C$_{0.12}$)$_{2}$ samples are displayed in Figs. 2(a) and (b).
Both samples consist of aggregated grains with size ranging from a few tenth to hundred $\mu$m.
In some of the intergrain regions, a minor secondary phase with darker contrast is clearly visible for (Mo$_{0.96}$Ti$_{0.04}$)$_{0.8}$B$_{2}$ but less evidence for (Mo$_{0.96}$Ti$_{0.04}$)$_{0.8}$(B$_{0.88}$C$_{0.12}$)$_{2}$.
EDX measurements indicate that this secondary phase is elemental boron in the former and elemental carbon in the latter, consistent with the XRD results shown above.
Figures 2(c-k) show the magnified SEM images (on a scale bar of 2.5 $\mu$m) and the corresponding EDX elemental maps taken on the large grains of these samples.
For both cases, a dense homogeneous bulk is observed and all the constituent elements are distributed uniformly.
In addition, the average Ti/(Mo+Ti) ratio is determined to be 0.04(1), in reasonable agreement with the nominal composition.
Nevertheless, the boron and carbon contents cannot be determined accurately by the EDX analysis due to their small atomic masses.
\begin{figure*}
\includegraphics*[width=14.8cm]{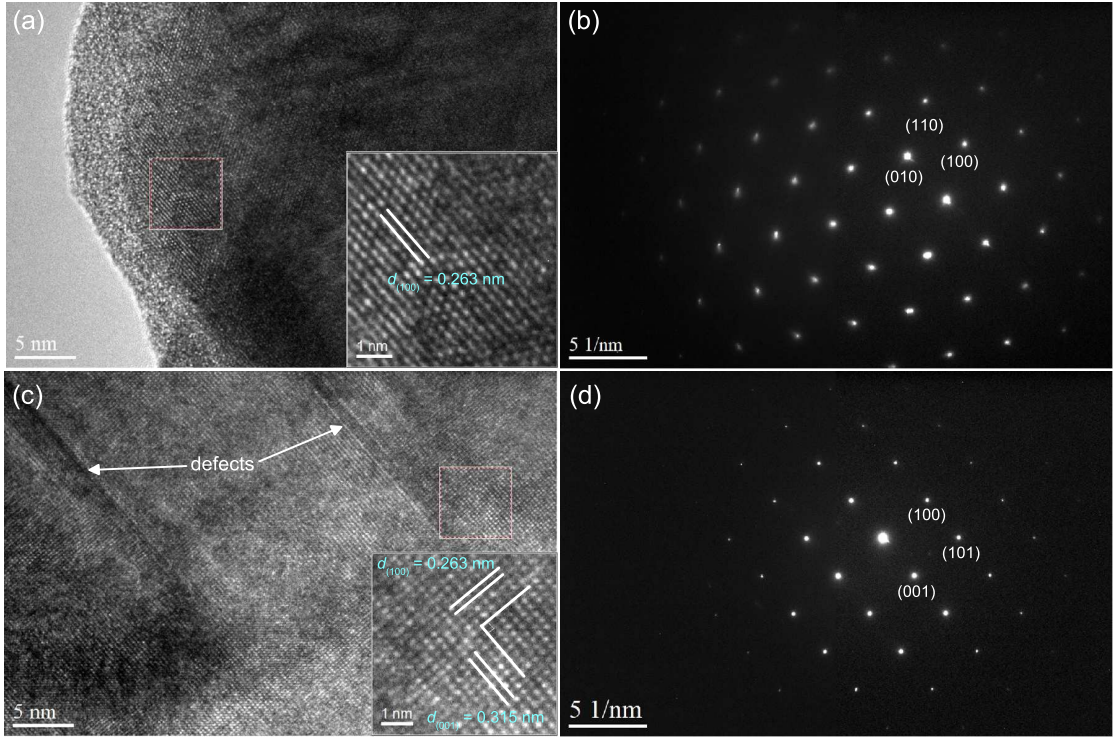}
\caption{
(a,b) High resolution TEM images and corresponding SAED pattern for the (Mo$_{0.96}$Ti$_{0.04}$)$_{0.8}$B$_{2}$ sample taken along the [0 0 1] zone axis.
(c,d) High resolution TEM images and corresponding SAED pattern for the same sample taken along the [0 1 0] zone axis.
In panel (c), the arrows mark the lattice defects.
}
\label{fig3}
\end{figure*}

\begin{figure}
\includegraphics*[width=14.8cm]{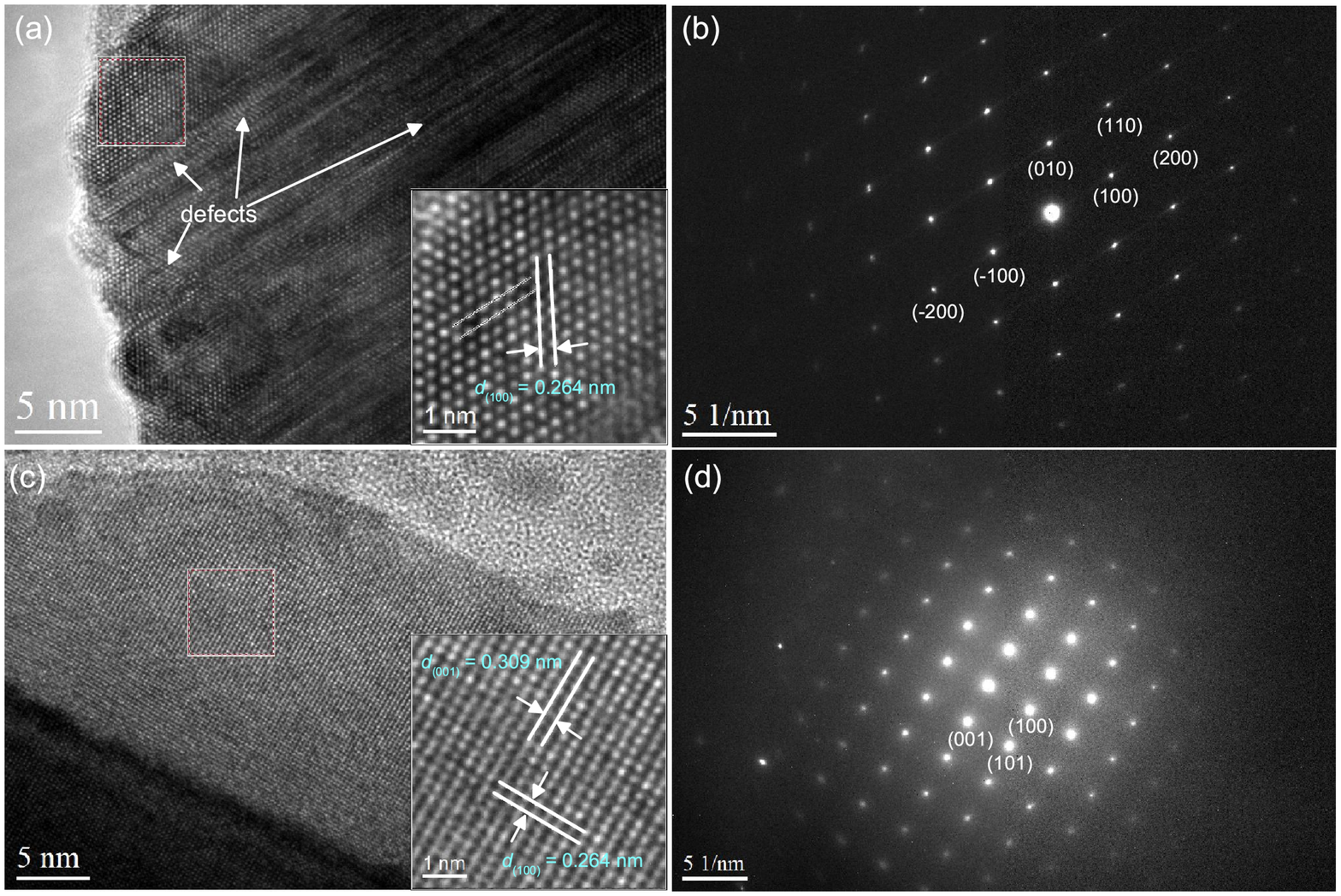}
\caption{
(a,b) High resolution TEM images and corresponding SAED patterns for the (Mo$_{0.96}$Ti$_{0.04}$)$_{0.8}$(B$_{0.88}$C$_{0.12}$)$_{2}$ sample taken along the [0 0 1] zone axis.
In panel (a), the arrows mark the lattice defects.
(c,d) High resolution TEM images and corresponding SAED patterns for the same sample taken along the [0 1 0] zone axis.
}
\label{fig4}
\end{figure}
Figure 3(a) shows the high-resolution TEM (HRTEM) image taken along the [0 0 1] zone axis for the (Mo$_{0.96}$Ti$_{0.04}$)$_{0.8}$B$_{2}$ sample.
One can see sharp lattice fringes with a spacing of 0.263 nm, which matches well with that between the (100) planes.
The corresponding selected area electron diffraction (SAED) displayed in Fig. 3(b) exhibits a well-defined spot pattern and the spots near the center can be indexed as the (100), (010) and (110) reflections.
The HRTEM image and corresponding SAED taken along the [0 1 0] zone axis for the same sample are displayed in Figs. 3(c) and (d).
Along this zone axis, two lattice spacings of 0.263 nm and 0.315 nm can be resolved, in line with those of the (100) and (001) planes, respectively.
Also, the spots near the center are indexable to the (100), (101) and (001) reflections.
It is worth noting that, similar to (Mo$_{0.96}$Zr$_{0.04}$)B$_{2}$ \cite{SC3}, planar defects along the $c$-axis are also detected for (Mo$_{0.96}$Ti$_{0.04}$)$_{0.8}$B$_{2}$, as indicated by the arrows in Fig. 3(c).
However, the absence of streaking in the SAED pattern [see Fig. 3(d)] implies that the density of such defects is significantly lower in the latter than in the former.
This is consistent with the expectation that the defects are suppressed by the presence of metal deficiency \cite{SC3}.

The HRTEM images and corresponding SAED patterns taken along the [0 0 1] and [0 1 0] zone axes for the (Mo$_{0.96}$Ti$_{0.04}$)$_{0.8}$(B$_{0.88}$C$_{0.12}$)$_{2}$ are shown in Figs. 4(a-d).
The lattice fringes remain well visible and the two resolved lattice spacings of 0.263 nm and 0.315 nm agree well with those of the (100) and (001) planes, respectively.
It is thus clear that the crystallinity dose not degrade upon carbon doping.
However, contrary to (Mo$_{0.96}$Ti$_{0.04}$)$_{0.8}$B$_{2}$, a number of planar defects are detected along the [0 0 1] zone axis of (Mo$_{0.96}$Ti$_{0.04}$)$_{0.8}$(B$_{0.88}$C$_{0.12}$)$_{2}$,
as seen in Fig. 4(a).
Indeed, the corresponding SAED pattern shown in Fig. 4(b) exhibits streaking along the (00$l$) directions.
Given that the streaks are absent along the [0 1 0] zone axis and the transition metal layers remain unchanged,
it is most likely that the planar defects come from the C-doped boron layers. In pristine boron layers, the boron atoms form a continuous network of six-membered rings.
The substitution of boron by smaller carbon is expected to distort the rings, which promotes the formation of planar defects.\\

\begin{figure}
\includegraphics*[width=10cm]{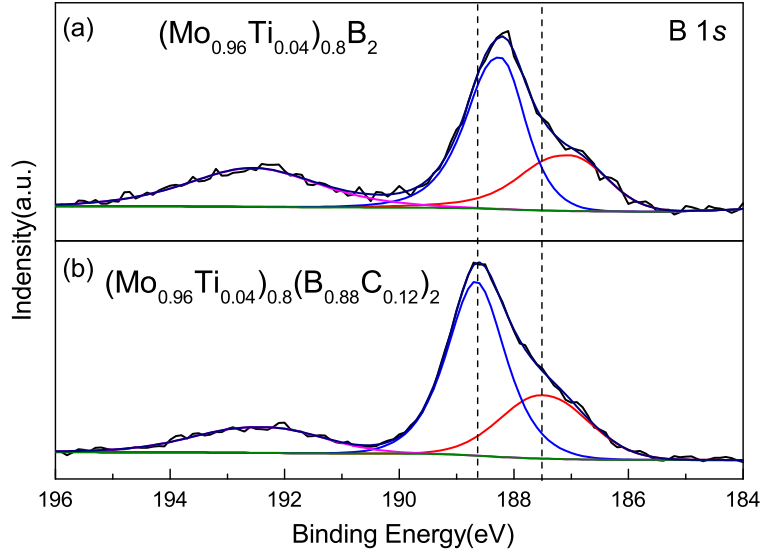}
\caption{
(a,b) B 1$s$ XPS spectrum and its deconvolution for the (Mo$_{0.96}$Ti$_{0.04}$)$_{0.8}$B$_{2}$ and (Mo$_{0.96}$Ti$_{0.04}$)$_{0.8}$(B$_{0.88}$C$_{0.12}$)$_{2}$ samples, respectively. The two vertical dashed lines are guides to the eyes.
}
\label{fig5}
\end{figure}
\noindent\textbf{3.3 XPS spectra}\\
Figures 5(a) and (b) show the B 1$s$ XPS spectra of the (Mo$_{0.96}$Ti$_{0.04}$)$_{0.8}$B$_{2}$ and (Mo$_{0.96}$Ti$_{0.04}$)$_{0.8}$(B$_{0.88}$C$_{0.12}$)$_{2}$ samples, respectively.
In the former case, the spectrum is deconvoluted into three peaks located at 187.1 eV, 188.2 eV and 192.6 eV, which can be assigned to B-B bonds \cite{B-Bbond}, B-Mo bonds and B-O bonds \cite{BO-Mobond}, respectively.
Similarly, the deconvolution of the spectrum of (Mo$_{0.96}$Ti$_{0.04}$)$_{0.8}$(B$_{0.88}$C$_{0.12}$)$_{2}$ gives three peaks located at 187.5 eV, 188.7 eV and 192.6 eV. Compared with (Mo$_{0.96}$Ti$_{0.04}$)$_{0.8}$B$_{2}$, the former two peaks shifts obviously towards higher binding energies while the latter one remains stationary. This suggests a contribution of B-C bonding since the the electronegtivity of carbon is larger than that of boron \cite{B-Cbond}. Note that this bonding inevitably involves the breaking of B-B bonds and induces a charge redistribution of the six-membered rings that favors the defect formation.\\

\noindent\textbf{3.4 Superconductiviting and normal-state properties}\\
\noindent\textbf{3.4.1 (Mo$_{0.96}$Ti$_{0.04}$)$_{0.8}$B$_{2}$}\\
\begin{figure}
\includegraphics*[width=14.8cm]{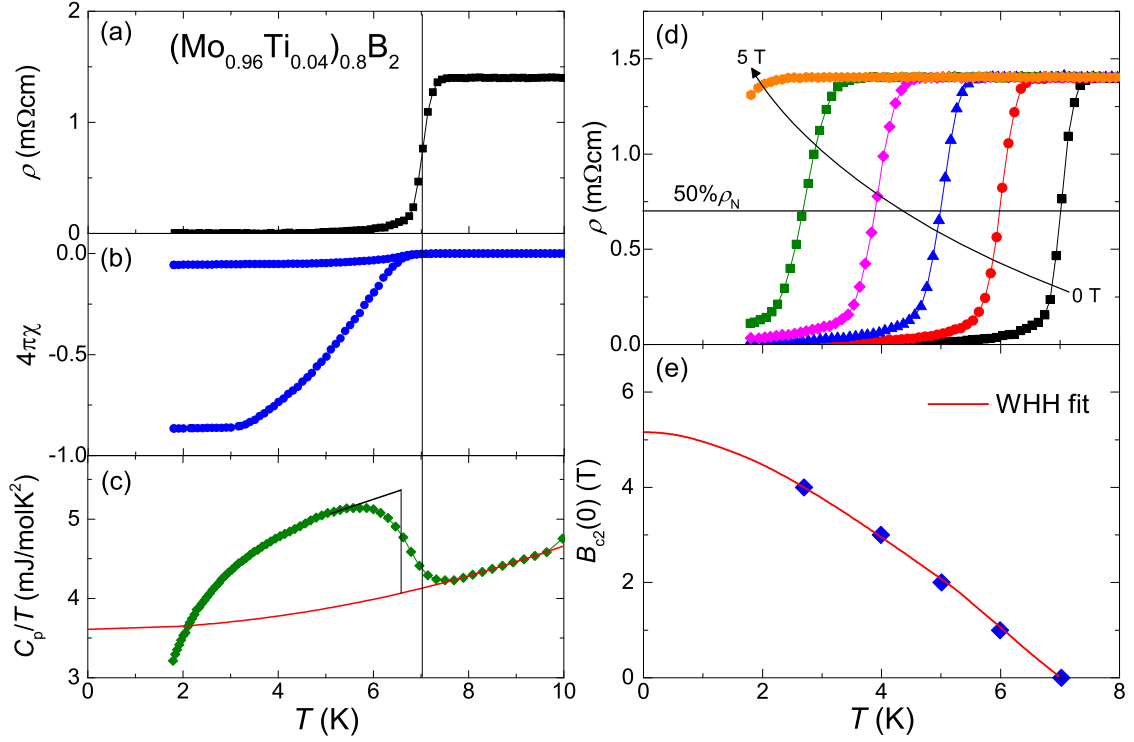}
\caption{
(a-c) Low temperature resistivity, magnetic susceptibility and specific heat, respectively, for the (Mo$_{0.96}$Ti$_{0.04}$)$_{0.8}$B$_{2}$ sample.
The vertical line is a guide to the eyes, showing the $T_{\rm c}$.
In panel (c), the black lines are entropy conserving construction to estimate the normalized specific-heat jump and the red line is a fit to the normal-state data by the Debye model.
(d) Temperature dependencies of resistivity under magnetic fields up to 5 T with a field increment of 1 T. The horizontal lines marks the level corresponding to half drop of the normal-state resistivity.(e) Upper critical field versus temperature phase diagram. The red line is a fit to the data by the WHH model.
}
\label{fig6}
\end{figure}
The low-temperature resistivity ($\rho$), magnetic susceptibility ($\chi$) and specific heat ($C_{\rm p}$) data for the (Mo$_{0.96}$Ti$_{0.04}$)$_{0.8}$B$_{2}$ sample are displayed in Figs. 6(a-c).
As can be seen, $\rho$ starts to drop on cooling below 7.4 K, indicating the onset of a superconducting transition.
The $\rho$ drop is sharp down to $\sim$6.8 K, but afterward becomes much slower, and achieves zero only below $\sim$6 K.
From the midpoint of the $\rho$ drop, the superconducting transition temperature $T_{\rm c}$ is determined to be 7.0 K.
Coinciding with $T_{\rm c}$, a diamagnetic transition in the zero-field cooling (ZFC) $\chi$ and a distinct $C_{\rm p}$ anomaly are detected.
In particular, the diamagnetic signal at 1.8 K corresponds to a shielding fraction of $-$4$\pi\chi$ $\approx$ 87.4\% , which is more than one order of magnitude larger than that reported previously \cite{SC3}.
These results indicate the bulk nature of superconductivity in (Mo$_{0.96}$Ti$_{0.04}$)$_{0.8}$B$_{2}$.

To extract the Sommerfeld coefficient $\gamma$, the normal-state $C_{\rm p}$ data is fitted by the Debye model,
\begin{equation}
C_{\rm p}/T = \gamma + \beta T^{2},
\end{equation}
where $\beta$ is the phonon specific heat coefficient.
This gives $\gamma$ = 3.61 mJ/molK$^{2}$ and $\beta$ = 0.01052 mJ/molK$^{4}$, and then the entropy-conserving construction yields the normalized specific heat jump $\Delta$$C_{\rm p}$/$\gamma$$T_{\rm c}$ = 0.41.
With $\beta$, the Debye temperature $\Theta_{\rm D}$ is calculated to be 802 K using the equation
\begin{equation}
\Theta_{\rm D} = (12\pi^{4} N R/5\beta)^{1/3},
\end{equation}
where $N$ = 2.8 is the number of atoms per formula unit and $R$ is the gas constant.
Then the electron-phonon coupling constant $\lambda_{\rm ep}$ is estimated to be 0.54 using the inverted McMillan formula \cite{Mcmillam}
\begin{equation}
\lambda_{\rm ep} = \frac{1.04 + \mu^{\ast} \rm ln(\Theta_{\rm D}/1.45\emph{T}_{\rm c})}{(1 - 0.62\mu^{\ast})\rm ln(\Theta_{\rm D}/1.45\emph{T}_{\rm c}) - 1.04},
\end{equation}
where $\mu^{\ast}$ = 0.13 is the Coulomb repulsion pseudopotential.
This, together with $\Delta$$C_{\rm p}$/$\gamma$$T_{\rm c}$, implies that (Mo$_{0.96}$Ti$_{0.04}$)$_{0.8}$B$_{2}$ is a weak coupling superconductor.

The upper critical field ($B_{\rm c2}$) of (Mo$_{0.96}$Ti$_{0.04}$)$_{0.8}$B$_{2}$ is determined by resistivity measurements under magnetic fields up to 5 T, the result of which is shown in Fig. 6(d).
The application of field leads to the shift of resistive transition toward lower temperatures.
For each field, $T_{\rm c}$ is determined using the same criterion at zero field, and the resulting temperature dependence of $B_{\rm c2}$ is plotted in Fig. 6(e).
Extrapolating the $B_{\rm c2}$($T$) data to 0 K using the Werthamer-Helfand-Hohenberg model \cite{WHH} gives the zero-temperature upper critical field $B_{\rm c2}$(0) = 5.2 T.
This $B_{\rm c2}$(0) is comparable to that of (Mo$_{1-x}$Sc$_{x}$)$_{1-\delta}$B$_{2}$ \cite{SC5} and much smaller than the Pauli paramagnetic limit $B_{\rm P}$(0) = 1.86$T_{\rm c}$ $\approx$ 13.1 T \cite{Paulilimit}, indicating that it is limited by the orbital effect.
Once $B_{\rm c2}$(0) is known, the Ginzburg-Landau (GL) coherence length $\xi_{\rm GL}$(0) is calculated to be 8.0 nm according to the equation
\begin{equation}
\xi_{\rm GL}(0) = \sqrt{\frac{\Phi_{0}}{2\pi B_{\rm c2}(0)}},
\end{equation}
where $\Phi_{0}$ = 2.07 $\times$ 10$^{-15}$ Wb is the flux quantum.\\
\begin{figure}
\includegraphics*[width=9.2cm]{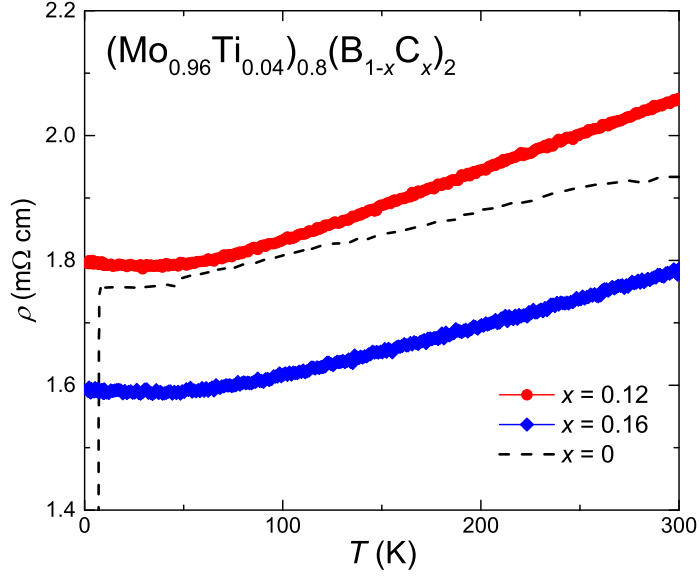}
\caption{
Temperature dependencies of resistivity up to 300 K for the (Mo$_{0.96}$Ti$_{0.04}$)$_{0.8}$(B$_{1-x}$C$_{x}$)$_{2}$ samples with $x$ = 0.12 and 0.16. The data for (Mo$_{0.96}$Ti$_{0.04}$)$_{0.8}$B$_{2}$ (dashed line) is also included for comparison.
}
\label{fig7}
\end{figure}

\noindent\textbf{3.4.2 (Mo$_{0.96}$Ti$_{0.04}$)$_{0.8}$(B$_{1-x}$C$_{x}$)$_{2}$}\\
Figure 7 shows the $\rho$($T$) curves for the (Mo$_{0.96}$Ti$_{0.04}$)$_{0.8}$(B$_{1-x}$C$_{x}$)$_{2}$ samples with $x$ = 0.12 and 0.16,
together with the data for (Mo$_{0.96}$Ti$_{0.04}$)$_{0.8}$B$_{2}$ (dashed line) for comparison.
For both C-doped samples, $\rho$ decreases smoothly with decreasing temperature, reflecting a metallic behavior. Compared with the C-free sample, the $\rho$ magnitude is larger for $x$ = 0.12 while smaller for $x$ = 0.16.
However, no $\rho$ drop is detected down to 1.8 K for the latter two cases, indicating that $T_{\rm c}$ is strongly suppressed for carbon content $x$ $\geq$ 0.12.\\

\noindent\textbf{3.4.3 Mechanism of $T_{\rm c}$ suppression by carbon doping}\\
The overall behavior of (Mo$_{0.96}$Ti$_{0.04}$)$_{0.8}$(B$_{1-x}$C$_{x}$)$_{2}$ is reminiscent of that observed in single crystalline Mg(B$_{1-x}$C$_{x}$)$_{2}$ \cite{MgB2C1}.
In the latter case, this suppression of $T_{\rm c}$ is mainly attributed to the effect of band filling since carbon is an electron dopant, which reduces the number of holes at the top of the boron $\sigma$ bands \cite{Tcsuppresion1}.
Contrary to MgB$_{2}$, the boron $\sigma$ orbitals are completely filled in MoB$_{2}$ \cite{SC3,MoB2bandstructure}. Instead, the electronic bands near the Fermi level ($E_{\rm F}$) are dominated by the Mo 4$d$ states and the boron states at the $E_{\rm F}$ are $\pi$ bonding in nature \cite{SC3}.
While this boron bonding is much weaker than that in MgB$_{2}$, its contribution to $\lambda_{\rm ep}$ is found to still play an important role in achieving the relatively high $T_{\rm c}$ in compressed MoB$_{2}$ \cite{MoB2bandstructure}.
It is thus reasonable to speculate that the boron $\pi$ states are also the key to superconductivity in (Mo$_{0.96}$Ti$_{0.04}$)$_{0.8}$B$_{2}$. The carbon substitution for boron denotes electrons, which fill the boron $\pi$ bands. This could lead to a reduction in electron-phonon coupling strength and consequently suppress the superconductivity in (Mo$_{0.96}$Ti$_{0.04}$)$_{0.8}$(B$_{1-x}$C$_{x}$)$_{2}$. Nonetheless, further studies in future are necessary to draw a more definitive conclusion.

\section{4. Conclusion}
In summary, we have investigated the structure and properties of (Mo$_{0.96}$Ti$_{0.04}$)$_{0.8}$(B$_{1-x}$C$_{x}$)$_{2}$ diborides prepared by the arc-melting method. The samples with $x$ = 0, 0.12, and 0.16 have a nearly single AlB$_{2}$-type phase with a uniform elemental distribution.
The carbon doping leads to a slight increase in the $a$-axis, a significant reduction in the $c$-axis, the formation of planar defects along the (100) planes, and a shift of the B 1$s$ peaks towards higher binding energies.
Moreover, the C-free (Mo$_{0.96}$Ti$_{0.04}$)$_{0.8}$B$_{2}$ ($x$ = 0) is confirmed to be a bulk superconductor below $T_{\rm c}$ = 7.0 K,
while no resistivity drop is observed down to 1.8 K for $x$ = 0.12 and 0.16. This suppression of superconductivity is attributed to weakening of electron-phonon coupling as a consequence of the electron filling of boron $\pi$ bands by carbon doping.
Our results not only represent the first study on the effect of carbon doping in transition metal diborides, but also help to better understand the superconductivity in this family of materials.
\section*{ACKNOWLEGEMENT}
We thank the foundation of Westlake University for financial support and the Service Center for Physical Sciences at Westlake University for technical assistance in SEM measurements.
The work at Zhejiang University is supported by the National Natural Science Foundation of China (12050003).

\end{document}